\documentclass[letterpaper, 10 pt, conference]{ieeeconf}  

\IEEEoverridecommandlockouts                              

\usepackage{graphicx}
\usepackage{cite}
\usepackage{amsmath,amsfonts,amssymb,array}
\usepackage{subcaption}
\usepackage{dblfloatfix}

\newtheorem{definition}{Definition}[section]
\newtheorem{proposition}{Proposition}[section]
\newtheorem{remark}{Remark}

\newcommand{\argmin}{\arg\!\min}
\newcommand{\argmax}{\arg\!\max}
\newcommand{\norm}[1]{\left\lVert#1\right\rVert}

\title{\LARGE \bf
Bayesian topology identification of linear dynamic networks}
\author{Shengling Shi, Giulio Bottegal and Paul M. J. Van den Hof 
\thanks{This project has received funding from the European Research Council (ERC), Advanced Research Grant SYSDYNET, under the European Union's Horizon 2020 research and innovation programme (grant agreement No 694504).}
\thanks{Shengling Shi, Giulio Bottegal and Paul M. J. Van den Hof are with the Department of Electrical Engineering,
        Eindhoven University of Technology, The Netherlands
        {\tt\small \{s.shi, g.bottegal, p.m.j.vandenhof\}@tue.nl}}%
}
\begin{document}

\maketitle
\thispagestyle{empty}
\pagestyle{empty}

\begin{abstract}
In networks of dynamic systems, one challenge is to identify the interconnection structure on the basis of measured signals. Inspired by a Bayesian approach in \cite{chiuso2012bayesian}, in this paper, we explore a Bayesian model selection method for identifying the connectivity of networks of transfer functions, without the need to estimate the dynamics. The algorithm employs a Bayesian measure and a forward-backward search algorithm. To obtain the Bayesian measure, the impulse responses of network modules are modeled as Gaussian processes and the hyperparameters are estimated by marginal likelihood maximization using the expectation-maximization algorithm. Numerical results demonstrate the effectiveness of this method.
\end{abstract}

\section{INTRODUCTION}
Estimation problems in system identification typically concern relatively simple structural setups, such as single-input-single-output or multipe-input-multiple-output, open-loop or closed-loop configurations \cite{ljung1987system}. Due to the increasing complexity of current technological systems, there is a need for  estimation techniques in large-scale interconnected dynamic systems, usually referred to as dynamic networks.


This work considers the network of transfer functions introduced in \cite{van2013identification}, where in the network, the nodes represent measured signals and the directed edges denote transfer functions, which are called modules. Identification problems in this setup involves multiple aspects, including estimation of one local module\cite{van2013identification} \cite{linder2017identification}, estimation of the topology \cite{materassi2010topological} \cite{chiuso2012bayesian}, estimation of the full network model \cite{weerts2018prediction} and identifiability aspects of the network models   \cite{hendrickx2018identifiability}. The network topology is sometimes assumed to be known in the estimation problems for dynamic networks \cite{dankers2016identification} \cite{gevers2015identification}. However, in many applications, estimation of the network topology is the main object of study, e.g. in systems biology \cite{hecker2009gene}, in social and political science \cite{lord2016inference} \cite{zhang2011causality}. 

Several methods using measures in the frequency domain can be found in \cite{materassi2010topological} \cite{materassi2012problem} \cite{shahrampour2015topology}. The approach in \cite{materassi2010topological} uses the coherence function and is built on the idea that nodes that are adjacent in a network should have a higher correlation than nodes that are more distant. However, this approach is developed for undirected tree structures only. A follow-up can be found in \cite{materassi2012problem}, where zero entries in a multivariate Wiener filter estimate of the dynamics are used to infer the topology. The approach in \cite{shahrampour2015topology} is formulated for state-space models, building on the observation that the inverse of the cross spectrum matrix changes if a subset of signals are set to zero.

Some approaches make use of regularized regression to enforce a subset of parameters belonging to the same module to zero; the topology is then identified by the remaining nonzero parameters. Typical regularization strategies exploit the $l_0$ norm penalty \cite{shahrampour2015topology} or the grouped version of the $l_1$ norm penalty  \cite{yuan2006model} \cite{bolstad2011causal}  on the parameter vector.

Search algorithms have also been employed to estimate the topology. An iterative algorithm known as \emph{block orthogonal matching pursuit} in compressed sensing employs a forward search procedure
\cite{eldar2009compressed}, while in the field of Bayesian networks, search algorithms coupled with Bayesian measures are commonly used to infer the topology \cite{chickering2002optimal}. However, the above approaches are not formulated for networks of transfer functions. A Bayesian approach formulated for dynamic networks can be found in \cite{chiuso2012bayesian}, where the impulse responses of the modules are modeled as Gaussian processes whose kernel is parameterized by hyperparameters; these hyperparameters are modeled as random variables whose probability density aims at enforcing the sparsity of the network.

Inspired by \cite{chiuso2012bayesian}, in this paper, a Bayesian model selection approach \cite{kass1995bayes} \cite{wasserman2000bayesian} is explored to solve the topology identification problem. While in \cite{chiuso2012bayesian} focus was on the joint estimation of topology and dynamics, our aim is to develop a Bayesian approach for topology identification, without estimating the dynamics. 

The approach in this work employs a Bayesian measure coupled with a forward-backward search algorithm to select the topology which optimizes the measure. To obtain the measure, a Gaussian prior distribution is assigned to the infinite impulse responses of the modules in dynamic networks. The hyperparameters of the prior are modeled as deterministic variables and estimated by maximizing marginal likelihood using a computationally attractive instance of the expectation-maximization (EM) algorithm; this constitute a major difference from the approach in \cite{chiuso2012bayesian}. In addition, comparing to that work, in this paper the topology is modeled as a random variable, which permits to incorporate structure prior information when required by specific applications.

\section{Problem formulation}
\label{sec:Problem}
The linear dynamic network model first introduced in \cite{van2013identification} is considered in this work:
\begin{equation}
w_j(t) = \sum_{i \in I \setminus j} G_{ji}(q) w_{i}(t) + H_{j}(q) e_j(t), \: j \in I, \label{eq:BasicModel}
\end{equation}
where $q^{-1}$ is the delay operator, i.e. $q^{-1}w_j(t)=w_j(t-1)$, $I = \{1, \cdots, L\}$ is the index set, $G_{ji}$ is a transfer operator and $e_j$ is a white noise process. The notation $w_I$ will be used to denote the set $\{w_j| j \in I\}$. With some abuse of notation, $w_j(t)$ denotes both a random variable and its realization. In addition, $Y \setminus B$ is used to denote the set difference between set $Y$ and $B$, i.e. $Y \setminus B = \{x \in Y | x \notin	 B\}$.

Combining (\ref{eq:BasicModel}) into a matrix form, the full model can be written as 
\begin{equation*}
w(t) = G(q) w(t) +H(q) e(t),
\end{equation*}
where $w(t) = [w_1(t),...,w_L(t)]^T$, $e(t) = [e_1(t),...,e_L(t)]^T$ and $H(q)$ is a diagonal matrix containing $H_j(q)$. The matrix $G(q)$ contains $G_{ji}(q)$ and has zero entries on its main diagonal.

The assumptions on (\ref{eq:BasicModel}) are summarized here:
\begin{itemize}
\item $w_j(t)$ can be measured for all $j$ and up to time $N$.

\item $(I-G(q))^{-1}$ is proper and stable.
	
\item $G_{ji}(q)$ is a stable and strictly proper rational transfer operator, $H_j(q)$ is monic and minimum-phase.

\item $e_j(t)$ is a white noise process and is also independent over nodes $j$. $e_j(t)$ follows a Gaussian distribution with an unknown standard deviation $\sigma_j$: 	$e_j(t) \sim \mathcal{N}(0,\sigma_j^2), \: \forall t.$

\end{itemize}

The topology of (\ref{eq:BasicModel}) can be defined as follows:
\begin{definition}
\label{def:topology}
The topology $\mathcal{G}$ corresponding to (\ref{eq:BasicModel}) is defined as $ \mathcal{G} = \{[i \: j]| G_{ji} \not= 0,\: i,\:j \in I\}$.
\end{definition}

The graphical representation of the topology is fully specified by $\mathcal{G}$, where the signals are represented by nodes and an directed edge $w_i \to w_j$ exists if $[i \: j] \in \mathcal{G}$. 
%
%
The problem of topology identification is to identify $\mathcal{G}$ of the data generating system given the measurements of $w_j(t)$ for all $t$ and all $j$. We shall denote such a set of measurements by $D$.


\section{Bayesian model selection}
\label{sec:BayeModelSelec}
To identify the topology, we need to define a measure that distinguishes two candidate structures on the basis of data. In this paper, a Bayesian model selection approach \cite{wasserman2000bayesian} is employed by modeling the topology as a random variable  and using measure $P(\mathcal{G}_1| D)/ P(\mathcal{G}_2| D)$ to compare two candidates, where $P(\mathcal{G}_i| D)$ is the posterior probability of $\mathcal{G}_i$ given data. The measure can be further formulated as
\begin{equation}
\frac{P(\mathcal{G}_1| D)}{P(\mathcal{G}_2| D)}  = \frac{P(D|\mathcal{G}_1)P(\mathcal{G}_1)}{P(D|\mathcal{G}_2)P(\mathcal{G}_2)} = \frac{P(D|\mathcal{G}_1)}{P(D|\mathcal{G}_2)}, \label{eq:BayesFactor}
\end{equation}
where $P(D|\mathcal{G})$ is the marginal likelihood and the second equality holds when there is no prior knowledge about the topology and thus $P(\mathcal{G}_i) = P(\mathcal{G}_j)$. In this work, we will assume that the second equality in  (\ref{eq:BayesFactor}) holds and for the reader who is interested in the structure prior, an example can be found in  \cite{werhli2007reconstructing}.  Thus, we will use $P(D|\mathcal{G}_1)/P(D|\mathcal{G}_2)$ which is also called \textit{Bayes factor} \cite{kass1995bayes}: taking the logarithm of $P(D|\mathcal{G})$, we can obtain an objective function whose maximization yields the topology with the highest marginal likelihood. Note that the Bayesian information criterion (BIC) is an approximation of $\log P(D|\mathcal{G})$ with a bounded error when $N \to \infty$ \cite{kass1995bayes}.

When the transfer operators are parameterized by a vector $\theta$, the marginal likelihood in (\ref{eq:BayesFactor}) can be obtained as
\begin{equation}
P(D|\mathcal{G}) = \int P(D|\theta, \mathcal{G}) P(\theta|\mathcal{G}) d\theta  \label{eq:MarginLikeCal},
\end{equation}
where $P(D|\theta, \mathcal{G})$ is the likelihood and $P(\theta|\mathcal{G})$ is the parameter prior distribution. Following the Bayesian approach, the topology maximizing $\log P(D|\mathcal{G})$ is the solution of the problem under study, which leads to the following problem:
\begin{equation}
\max_{\mathcal{G} \in \mathcal{G}_{\text{set}}} \log P(D|\mathcal{G}), \label{eq:OptimizationProblem}
\end{equation}
where $\mathcal{G}_{\text{set}}$ denotes the set of all possible graphs.
To solve (\ref{eq:OptimizationProblem}), we need to address i) the choice of $P(\theta|\mathcal{G})$, ii) the calculation of the integration in (\ref{eq:MarginLikeCal}), and iii) the solver to select the topology when there are a large number of candidates. Theses issues are discussed in the next section.

\section{Bayesian topology identification}
\label{sec:MainResults}
\subsection{Reformulation of the problem}
Model (\ref{eq:BasicModel}) can be reformulated as
\begin{equation}
w_j(t) = \hat{w}_j(t|t-1) + e_j(t), \label{eq:PredictorModel}
\end{equation}
where $\hat{w}_j(t|t-1)$ is the one-step ahead predictor, namely
\begin{equation*}
\hat{w}_j(t|t-1) = [1- H^{-1}_j(q)] w_j(t) + \sum_{i \in I \setminus j} \frac{G_{ji}(q)}{H_j(q)} w_i(t),
\end{equation*}
and $\frac{G_{ji}(q)}{H_j(q)} = \sum_{k=1}^\infty \theta_{ji,k} q^{-k}, \: 1- H^{-1}_j(q) = \sum_{k=1}^\infty \theta_{jj,k} q^{-k}$,
under the assumptions that $G_{ji}(z)$ is stable and $H_j(z)$ is minimum-phase \cite{ljung1987system}.
Each infinite-order impulse response is approximated by a finite order $n$, which will have no impact on the performance of the method when $n$ is sufficiently large. Then the compact form of (\ref{eq:PredictorModel}) containing measurements up to time $N$ can be written as
\begin{equation}
w^N_j = \sum_{i\in I} A_{ji} \theta_{ji} + e^N_j, \quad j \in I \label{eq:PredictorMISO}
\end{equation}
where $w^N_j = [w_j(1),..., w_j(N)]^T$, $\theta_{ji}= [\theta_{ji,1},..., \theta_{ji,n}]^T$, $e^N_j = [e_j(1),..., e_j(N) ]^T$,  and $A_{ji}$ is a Toeplitz matrix containing the measurements of $w_i(t)$.
Equation (\ref{eq:PredictorMISO}) can also be written as $w^N_j = A_{j} \theta_{j} + e^N_j$, where $A_j=[A_{j1},...,A_{jL}]$ and $\theta_{j} = [\theta_{j1}^T,...,\theta_{jL}^T]^T$. Equivalently, the problem considered in this work can be also formulated based on (\ref{eq:PredictorMISO}) as the identification of the set $\bar{\mathcal{G}} = \{[i \: j]| \theta_{ji} \not= 0,\: i,\: j \in I \}$.
Note that $\bar{\mathcal{G}}$ is defined on the predictor model (\ref{eq:PredictorModel}) while $\mathcal{G}$ is defined on (\ref{eq:BasicModel}). It can be found that $\mathcal{G}$ is equivalent to $\bar{\mathcal{G}}$ when the self-loops in $\bar{\mathcal{G}}$ are removed. Even if the algorithm is designed to recover $\bar{\mathcal{G}}$, the notation $\mathcal{G}$ is still used in place of $\bar{\mathcal{G}}$ and the self-loops are made implicit to improve the readability.

\subsection{Decomposition of the objective function}
In this section, we show that the objective function $\log P(D|\mathcal{G})$  can be decomposed into a set of independent terms corresponding to MISO problems, where each MISO topology identification problem can be solved independently.


Based on (\ref{eq:MarginLikeCal}), it can be seen that $P(D|\mathcal{G})$ can be factorized by decomposing $P(D|\theta, \mathcal{G})$ and $P(\theta|\mathcal{G})$. Due to the Bayes' rule and
the assumption that the noises are white and independent over nodes, if each MISO model is independently parameterized, it holds that the likelihood can be factorized as
\begin{equation}
P(D|\theta, \mathcal{G}) = \prod_{j=1}^L \prod_{t=1}^N P(w_j(t)| \hat{w}_j(t|t-1) ). \label{eq:MISOlike}
\end{equation}
The independent parameter assumption implies that the term $P(\theta| \mathcal{G})$ in (\ref{eq:MarginLikeCal}) satisfies
\begin{equation}
P(\theta| \mathcal{G})=\prod_{j=1}^L P(\theta_j|\mathcal{G}_j), \label{eq:ParaIndep2}
\end{equation}
where $\mathcal{G}_j$ and $\theta_j$ denote the topology and the parameter vector of one MISO model, respectively.
Thus, given (\ref{eq:BasicModel}) and the parameter independence assumption, the marginal likelihood in (\ref{eq:OptimizationProblem}) can be decomposed into $L$ independent terms as $\log P(D|\mathcal{G}) = \sum_{j=1}^L \log P(D_j | \mathcal{G}_j)$,
where $D_j$ denotes the data relevant to a single MISO problem of the type (\ref{eq:PredictorMISO}) and
\begin{equation}
\log P(D_j | \mathcal{G}_j)
\triangleq \log \int \prod_{t=1}^N P(w_j(t)| \hat{w}_j(t|t-1)) P(\theta_j|\mathcal{G}_j) d\theta_j. \label{eq:MISOml}
\end{equation}
Since each term is a function of the MISO topology, the search algorithm for the MISO topology can then be parallelized to obtain the overall network topology.

\subsection{Objective function: Parameter prior and integration}

Due to the independence among the MISO problems, in this section we describe the developed algorithm for a single MISO model of the type (\ref{eq:PredictorMISO}).


Firstly, we need to specify the dependence of $P(D_j|\theta_j, \mathcal{G}_j)$ and $P(\theta_j|\mathcal{G}_j)$ on one particular structure $\mathcal{G}_j$. Given one topology $\mathcal{G}_j = \{[i_1 \: j],..., [i_p \: j]\}$, $P(\theta_j|\mathcal{G}_j)$ considers the distribution of the parameter vector formulated based on $\mathcal{G}_j$, i.e. $ \theta_j|\mathcal{G}_j = \begin{bmatrix} \theta_{j i_1}^T & \cdots & \theta_{j i_p}^T \end{bmatrix}^T$. Note that with some abuse of notation, $\theta_j|\mathcal{G}_j$ denotes a vector formulated based on the indexes in $\mathcal{G}_j$. In addition, the likelihood function $P(D_j|\theta_j, \mathcal{G}_j)$ is calculated based on the model $w^N_j = (A_j| \mathcal{G}_j) \times (\theta_j | \mathcal{G}_j) + e^N_j$, where $A_j| \mathcal{G}_j = \begin{bmatrix} A_{j i_1} & \cdots & A_{j i_p} \end{bmatrix}$.

\subsubsection*{\textbf{Parameter prior}}
Following the kernel-based approach for system identification \cite{pillonetto2010new}, since the prior knowledge that the impulse responses should decay with time is available, the parameter prior $P(\theta_j |\mathcal{G}_j)$ is chosen from \cite{pillonetto2014kernel} as
\begin{equation}
\theta_j|\mathcal{G}_j  \sim \mathcal{N}(0, K_j) \label{eq:parameter prior},
\end{equation}
where $K_j$ is a block diagonal matrix as $K_j= diag(\lambda_{ji_1} \bar{K}(\beta_{ji_1}),\cdots,\lambda_{ji_p} \bar{K}(\beta_{ji_p}))$ ,
$\bar{K}(\beta_{ji})$ is a $n \times n$ matrix and the $(k,q)$ entry of $\bar{K}(\beta_{ji})$ is defined by $\beta_{ji}^{\max(k,q)}$. It is required that $\lambda_{ji} > 0$ and $\beta_{ji} \in [0 \; 1)$. For this choice of kernel $\bar{K}$, $\beta_{ji}$ regulates the velocity of the decay of the impulse responses. Therefore, the module priors depend on the unknown hyperparameter vectors, i.e. $\lambda_j | \mathcal{G}_j= \begin{bmatrix} \lambda_{ji_1} & \cdots & \lambda_{ji_p} \end{bmatrix}^T$ and $\beta_j | \mathcal{G} = \begin{bmatrix} \beta_{ji_1} & \cdots & \beta_{ji_p} \end{bmatrix}^T$. Since every MISO problem will be assigned an independent parameter prior as (\ref{eq:parameter prior}), equation~(\ref{eq:ParaIndep2}) is satisfied.

\subsubsection*{\textbf{Integration}}
Denote $\eta_j= \begin{bmatrix} \sigma_j & \lambda_j^T & \beta_j^T \end{bmatrix}^T$, where the dependencies of $\lambda_j$ and $\beta_j$ on $\mathcal{G}_j$ are implicit. Based on (\ref{eq:PredictorMISO}) and (\ref{eq:parameter prior}), given one particular $\mathcal{G}_j$, (\ref{eq:MISOml}) can be obtained in a closed form. After scaling and removing a constant term, we can obtain that
\begin{align}
& J(\mathcal{G}_j; \eta_j) = 2 \log P(D_j|\mathcal{G}_j; \eta_j) - \text{constant term} \nonumber \\
= & - (w^N_j)^T \Gamma_j^{-1}w^N_j - \log \det\Gamma_j, \label{eq:objective}
\end{align}
where $\Gamma_j = \sigma_j^{2} I_N + A_jK_jA_j^T$ and the dependencies of $A_j$ and $K_j$ on a particular topology $\mathcal{G}_j$ are implicit. Note that $\Gamma_j$ is also a function of $\eta_j$. Since $\eta_j$ is unknown, an estimate of $\eta_j$ has to be computed first and then we can use $J(\mathcal{G}_j; \hat{\eta}_j)$ as the objective function for the topology estimation problem.

\subsubsection*{\textbf{Estimation of hyperparameters}}
To obtain an estimate of $\hat{\eta}_j$, we estimate the hyperparameter vector associated to the full graph, namely $\eta^{full}_j$. Then, given a graph $\mathcal{G}_j$, the corresponding hyperparameter vector $\hat{\eta}_j$ associated to that graph can be obtained by neglecting those hyperparameters associated to zero modules (i.e., missing edges in the graph). This procedure avoids the re-estimation of $\eta_j$ for all different graphs and reduces the computational cost. The hyperparameter vector $\eta^{full}_j$ is estimated by solving the following marginal likelihood problem:
\begin{equation}
\hat{\eta}^{full}_j = \argmax_{\eta^{full}_j} \log P(D_j|\mathcal{G}^{full}_{j}; \eta^{full}_j), \label{eq:EMproblem}
\end{equation}
where $\mathcal{G}^{full}_{j}$ is a full graph, i.e. $\mathcal{G}^{full}_{j}=\{[1 \: j],...,[L \: j]\}$. A local optimum of this problem can be found by the EM algorithm \cite{bottegal2016robust}.

Assuming that an estimate $\hat{\eta}_j^{(k)}$ of $\eta^{full}_j$ is available at the $k$-th iteration of the EM algorithm, an update estimate is obtained by the following steps:

(\textbf{E-step}) Compute
\begin{equation}
Q(\eta_j,\hat{\eta}_j^{(k)}) = E_{P(\theta_j|w^N_j;\hat{\eta}_j^{(k)})}[ \log P(\theta_j,w^N_j; \eta_j)]; \label{eq:Estep}
\end{equation}

(\textbf{M-step}) Compute
\begin{equation}
\hat{\eta}_j^{(k+1)} = \argmax_{\eta_j \in \mathcal{V}} Q(\eta_j,\hat{\eta}_j^{(k)}). \label{eq:Mstep}
\end{equation}
Note that for a MISO problem, the input and the graph are regarded as fixed and thus implicit in (\ref{eq:Estep}).

\begin{proposition} \label{pro:EM}
Denote $\hat{\eta}^{(k)}$ as the estimate of the hyperparameter vector at the $k$th iteration of the EM algorithm used to solve (\ref{eq:EMproblem}). Then, according to (\ref{eq:Estep}) and (\ref{eq:Mstep}), $\hat{\eta}^{(k+1)}$ is obtained with the following update rules:
\begin{itemize}
\item The hyperparameter $\hat{\sigma}^{k+1}_j$ is obtained as
\begin{equation}
\hat{\sigma}^{k+1}_j = \sqrt{\frac{M^{(k)}}{N}}, \label{eq:EM_sigma}
\end{equation}
where
\begin{align*}
M^{(k)} =& (w^N_j)^T w^N_j - 2(w^N_j)^T A_j \hat{C}_j^{(k)}w^N_j \\
& + tr[A_j^TA_j \hat{\Delta}_j^{(k)}],\\
\hat{C}_j^{(k)} =& [\hat{\sigma}_j^{(k)}]^{-2} [\hat{\Sigma}_j^{(k)}] ^{-1}A_j^T,\\
\hat{\Sigma}_j^{(k)} =& [\hat{\sigma}_j^{(k)}]^{-2} A_j^T A_j + [K_j(\hat{\lambda}_j^{(k)},\hat{\beta}_j^{(k)})]^{-1},\\
\hat{\Delta}_j^{(k)} =& [\hat{\Sigma}_j^{(k)}]^{-1}+\hat{C}_j^{(k)} w^N_j (w^N_j)^T [\hat{C}_j^{(k)}]^T.
\end{align*}

\item The hyperparameter $\hat{\beta}^{k+1}_{ji}$, $i=1,...,L$, is obtained as
\begin{align}
\hat{\beta}^{k+1}_{ji} = \argmin_{\beta_{ji} \in [0\:1)} & n \log [tr(\bar{K}^{-1}(\beta_{ji}) \hat{\Delta}^{(k)}_j[i])] \nonumber \\
 & + \log \det \bar{K}(\beta_{ji}), \label{eq:EM_beta}
\end{align}
where $\hat{\Delta}^{(k)}_j[i]$ is a square sub-matrix obtained from $\hat{\Delta}^{(k)}_j$ by the $[(i-1)n+1]$-th row and column until the $(in)$-th row and column of $\hat{\Delta}^{(k)}_j$.

\item The hyperparameter $\hat{\lambda}^{k+1}_{ji}$, $i=1,...,L$, is obtained as
\begin{equation}
\hat{\lambda}^{k+1}_{ji} = \frac{1}{n} tr[\bar{K}^{-1}(\hat{\beta}^{k+1}_{ji}) \hat{\Delta}^{(k)}_j[i]]. \label{eq:EM_lamda}
\end{equation}
\end{itemize}
\end{proposition}

It can be found that (\ref{eq:EMproblem}) is decomposed into a set of optimization problems with scalar optimization variables for estimating $\beta$ and closed-form solutions for estimating $\sigma$ and $\lambda$. The computational speed of the above algorithm can be further improved by exploiting the factorization of $\bar{K}$ \cite{carli2014maximum} \cite{bottegal2016robust}, which is also implemented in the algorithm.
\subsection{Algorithm for optimization}
The objective function of problem (\ref{eq:OptimizationProblem}) has been formulated in (\ref{eq:objective}), where $J(\mathcal{G}_j; \hat{\eta}_j)$  is used to replace $\log P(D_j | \mathcal{G}_j)$  and $\hat{\eta}_j$ is obtained as $\hat{\eta}_j = \hat{\eta}^{full}_j| \mathcal{G}_j$. The next step is to design the solver for the optimization problem.

Since the number of all possible directed graphs in $\mathcal{G}_{\text{set}}$ is $2^{L^2-L}$, it is infeasible to consider all the candidates. Following \cite{chickering2002optimal}, a forward-backward greedy search algorithm is implemented to find a local optimum of (\ref{eq:OptimizationProblem}). 
Recall that the graph of the predictor model is considered here, so that self-loops are generally present. The algorithm initializes a graph with only self-loops and then starts the edge-addition phase, where at each iteration, the edge which most improves the objective value is added to the graph from the previous iteration. The iterations stop when no improvement can be found by adding edges.

Given the final graph of the edge-addition phase, the algorithm starts the edge-deletion phase, where at each iteration, one edge is removed from the graph of the previous iteration if such deletion improves the objective function comparing to the removal of other edges. The final output of the algorithm is obtained when no improvement in the objective value can be found by deleting any edge.

%

As mentioned earlier, due to the decomposition in (\ref{eq:MISOml}), the search algorithm can be applied to every MISO problem separately, merging the outcomes to obtain the network topology.

\subsection{Final algorithm}
After the formulation of the objective function and the greedy search algorithm, the algorithm is now complete and summarized in this section. Firstly, recall that $\hat{\eta}^{full}_j$ obtained in the previous step is for a full graph and thus, given a structure $\mathcal{G}_j$, $\hat{\eta}_j$ should be reformulated as $\hat{\eta}^{full}_j |\mathcal{G}_j = \begin{bmatrix}  \hat{\sigma}_j & (\hat{\lambda}^{full}_j | \mathcal{G}_j) ^T & (\hat{\beta}^{full}_j | \mathcal{G}_j)^T \end{bmatrix}^T$.
To simplify the notation, the index $j$ is dropped in the algorithm.

\textbf{Algorithm} (BS - Bayesian Search): Inputs: data $D$; Outputs: $\hat{\mathcal{G}}$
\begin{enumerate}
 \item Obtain $\hat{\eta} = \max_{\eta} \log p(D|\mathcal{G}_{full}; \eta)$ by EM algorithm
 \item Initialize $\mathcal{G}^{(0)} = \{[j \:  j]\}$ and $Edge = \{[1 \: j], \cdots, [L \: j]\}$
 \item  For $b = 1:L-1$ (Edge-addition phase)
 \begin{itemize}
 	\item $Edge = Edge \setminus \hat{\mathcal{G}}^{(b-1)}$
 	\item $[\hat{i}\: j]= arg\max_{[i \: j] \in Edge} J(\{\hat{\mathcal{G}}^{(b-1)}, \: [i \: j]\}; \hat{\eta}  )$
 	\item if  $J(\{\hat{\mathcal{G}}^{(b-1)}, \: [\hat{i}  \: j]   \}; \hat{\eta}  ) - J(\hat{\mathcal{G}}^{(b-1)}; \hat{\eta} ) > \tau$
 	\begin{itemize}
 		\item[$\diamond$] $\hat{\mathcal{G}}^{(b)} =  \{\hat{\mathcal{G}}^{(b-1)}, \: [\hat{i}  \: j]  \}$
 	\end{itemize}
 	\item else
    \begin{itemize}
 		\item[$\diamond$]  break loop
 	\end{itemize}
 \end{itemize}
 \item Initialize for the second phase: $\hat{\mathcal{G}}^{(0)} = \hat{\mathcal{G}}_{FinalAddition}$
 \item For $d = 1:| \hat{\mathcal{G}}^{(0)}  |$ (Edge-deletion phase)
 \begin{itemize}
	 \item $[\hat{i} \: j] = arg\max_{[i\: j] \in \hat{\mathcal{G}}^{(d-1)}} J( \hat{\mathcal{G}}^{(d-1)}\setminus [i \: j]; \hat{\eta}  )$
     \item if  $J( \hat{\mathcal{G}}^{(d-1)}\setminus[\hat{i} \: j]; \hat{\eta} )- J(\hat{\mathcal{G}}^{(d-1)}; \hat{\eta} ) > \tau$
 	 \begin{itemize}
 		\item[$\diamond$] $\hat{\mathcal{G}}^{(d)} =  \hat{\mathcal{G}}^{(d-1)}\setminus[\hat{i} \: j]$
 	 \end{itemize}
 	\item else
    \begin{itemize}
 		\item[$\diamond$]  break loop
 	\end{itemize}
 \end{itemize}
\end{enumerate}
The tolerance $\tau$, determining whether an edge should be added or removed, is chosen to be zero as  default value; its suggested range is $[0,10]$, see \cite{kass1995bayes}. 
\begin{remark}
To empirically validate the choice of using the estimate of $\hat \eta^{full}$ under the full graph, the BS algorithm is compared with its variant using an iterative EM approach, which re-estimates $\hat{\eta}$ by the EM algorithm under every iteration of the search algorithm. We call this procedure the iterative-EM BS algorithm. \QED
\end{remark}

Comparing to the approach in \cite{chiuso2012bayesian}, the main difference of the BS algorithm is that the hyperparameters are modeled as deterministic variables and then estimated by the EM algorithm. By contrast, in \cite{chiuso2012bayesian}, the hyperparameters are modeled as random variables and a prior distribution of the hyperparameters is also used. The choice of modeling also the hyperparameters as random variables requires designing their prior distribution, which usually requires to include additional hyper-hyperparameters that may be difficult estimate.

\section{Kernel-based group Lasso}
\label{sec:lasso}
The performance of the BS algorithm is compared with the group Lasso (GLasso) estimator \cite{yuan2006model}, which is formulated on the basis of (\ref{eq:PredictorMISO}) as
\begin{equation}
\min_{\theta_j} \frac{1}{2} \norm{w^N_j - A_j \theta_j}_2^2 + \delta_j \sum_{i=1}^L \norm{\theta_{ji}}_2. \label{eq:Glasso}
\end{equation}
Here, the topology estimation problem is also divided into independent MISO problems. It is also of interest to see if the performance of (\ref{eq:Glasso}) can be improved by incorporating the covariance matrix in (\ref{eq:parameter prior}) into the regularization term. This kernel-based GLasso can be formulated as
\begin{equation}
\min_{\theta_j} \frac{1}{2} \norm{w^N_j - A_j \theta_j}_2^2 + \delta_j \sum_{i=1}^L \sqrt{ \theta_{ji}^T \bar{K}(\beta_j,n)^{-1} \theta_{ji}}. \label{eq:KernelGlasso}
\end{equation}
To reduce the computational complexity, we choose to have the same hyperparameters $\beta_j$ for all  modules of each MISO problem. To select $\delta_j$ and $\beta_j$, cross validation can be employed. After having the estimated parameters, the topology can be obtained by checking if the $l_2$ norm of the parameter vector corresponding to one module is zero.
\section{Numerical results}
\label{sec:Numerical}

\begin{figure*}
\centering
\begin{subfigure}{.68\columnwidth}
\includegraphics[width=\columnwidth]{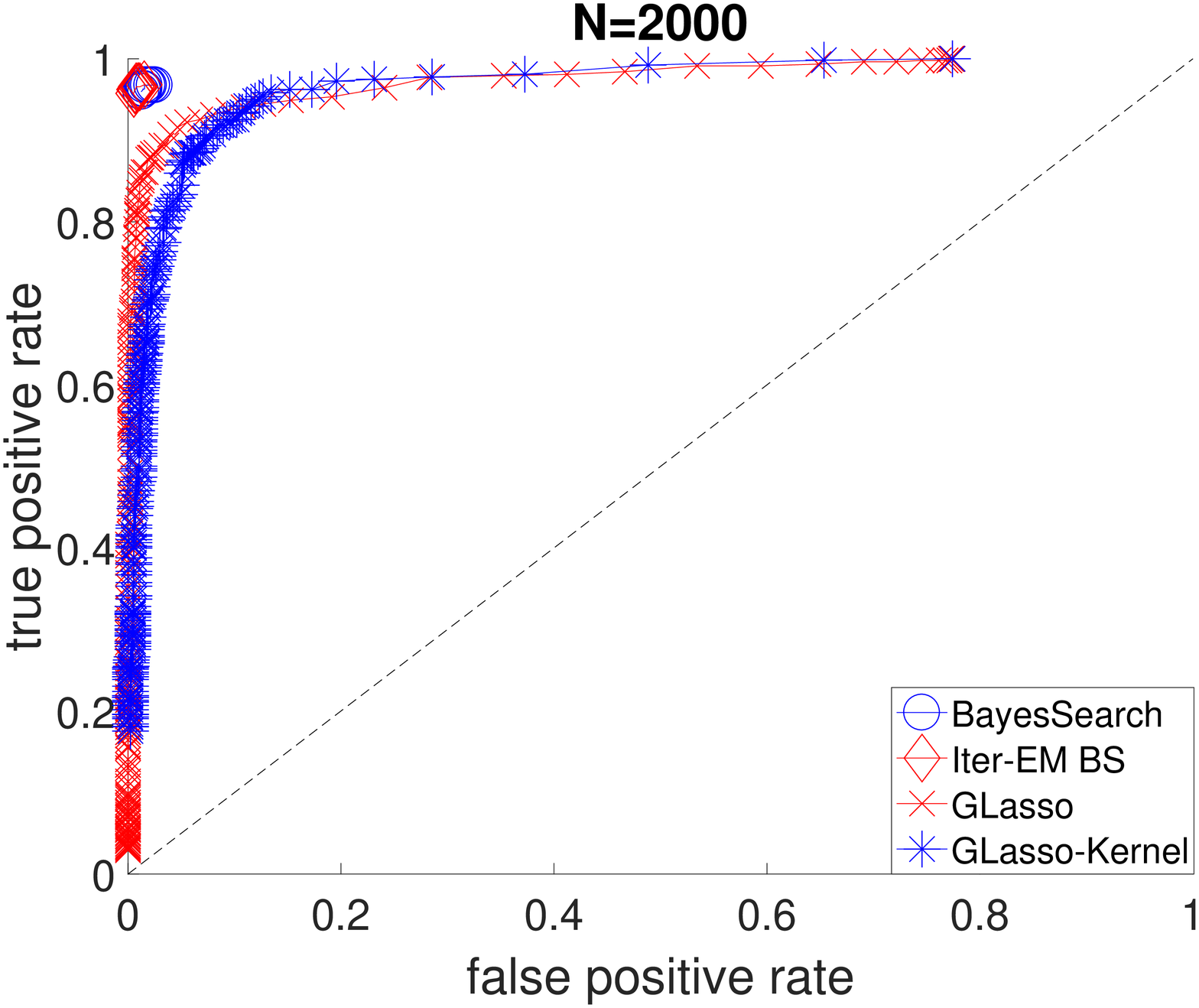}%
\caption{}%
\label{fig:ROC1}%
\end{subfigure}\hfill%
\begin{subfigure}{.68\columnwidth}
\includegraphics[width=\columnwidth]{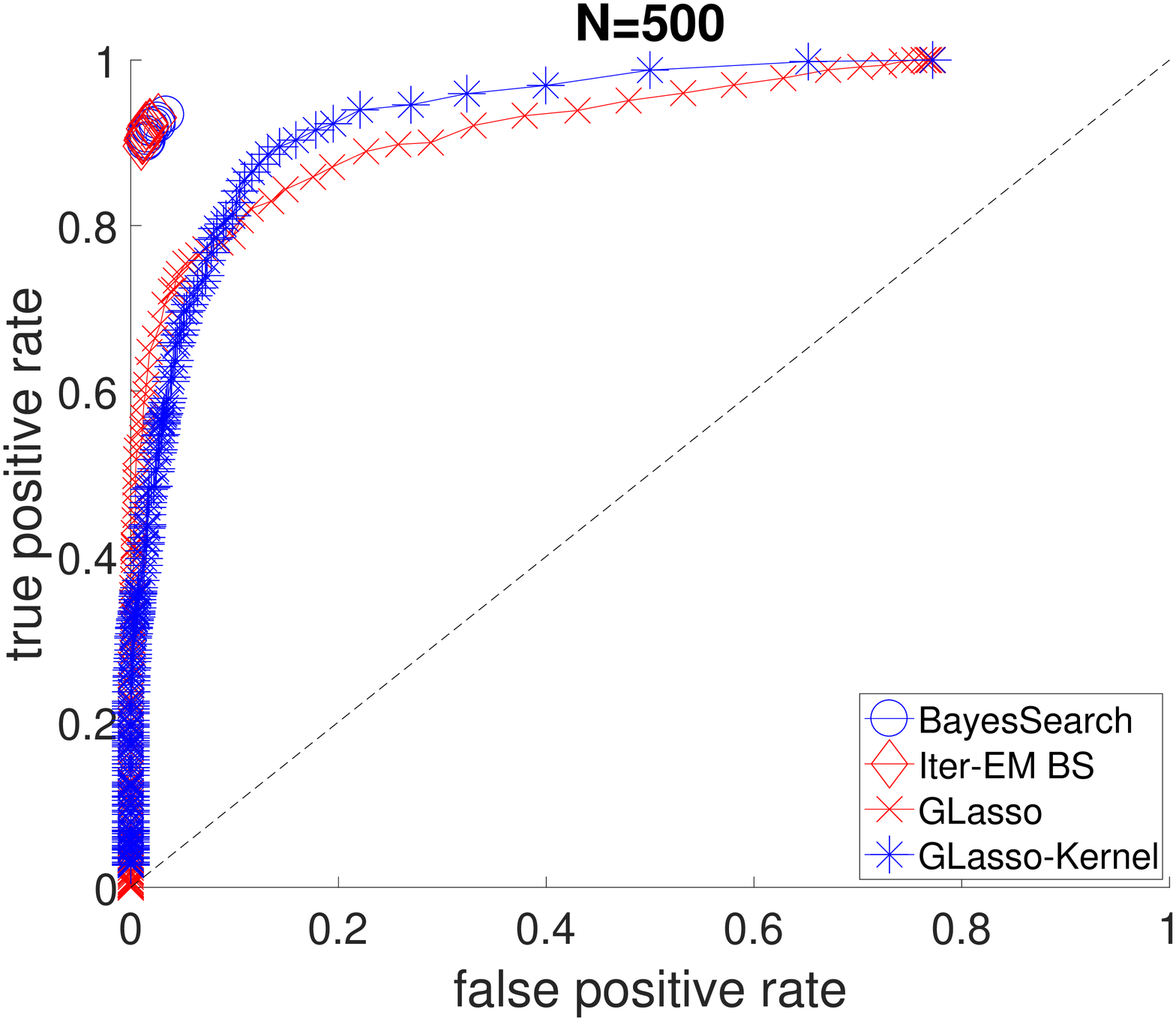}%
\caption{}%
\label{fig:ROC15}%
\end{subfigure}\hfill%
\begin{subfigure}{.68\columnwidth}
\includegraphics[width=\columnwidth]{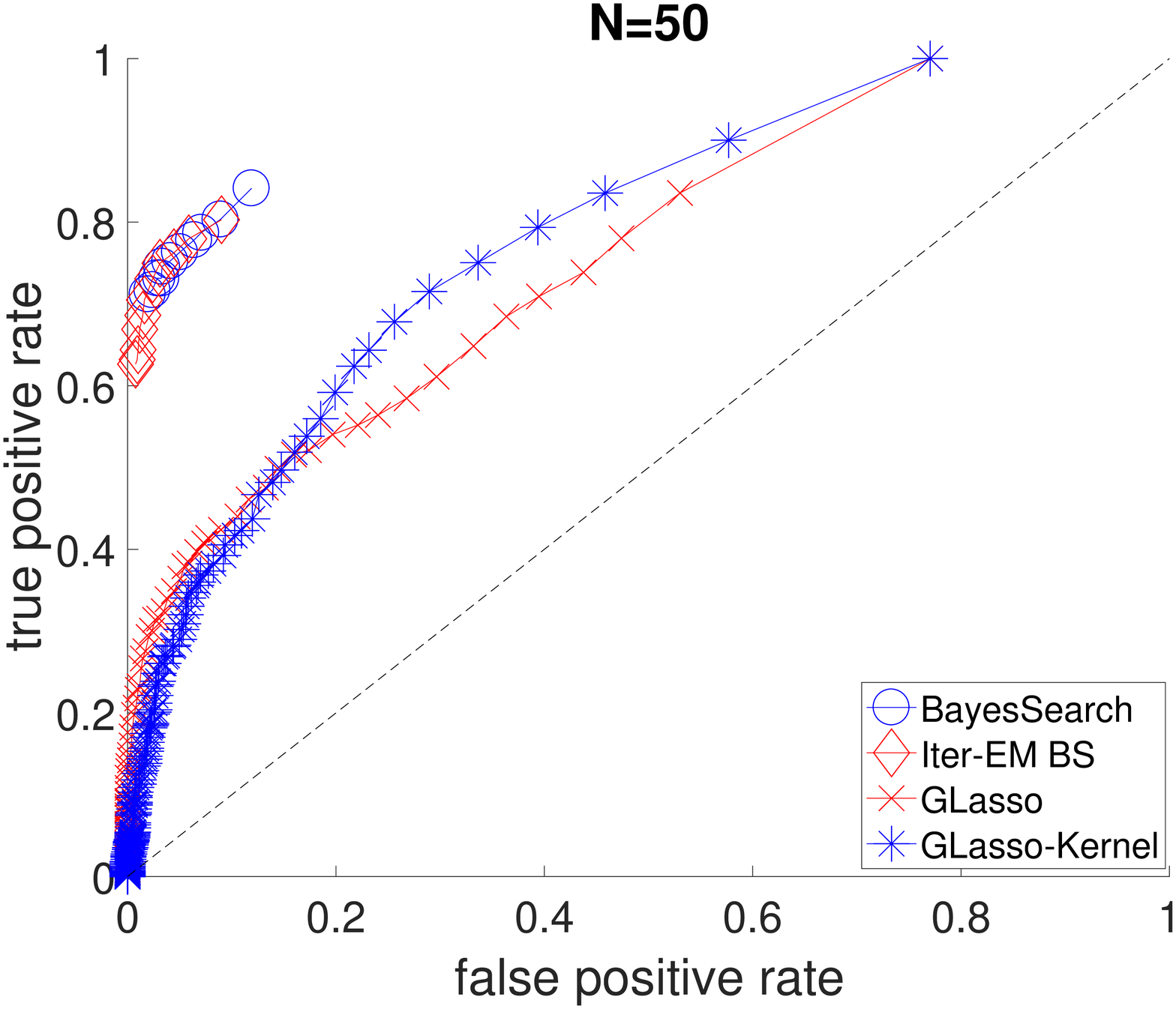}%
\caption{}%
\label{fig:ROC2}%
\end{subfigure}%
\caption{\small{$TPR$ vs $FPR$ over tuning parameters for different data length: $N = 2000$ (left), $N=500$ (middle), $N=50$ (right)}}
\label{figabc}
\end{figure*}

To evaluate the performance of the algorithms, an existing edge in the network is labeled as one positive instance; its absence is labeled as one negative instance. Let $P$ denote the total number of positives and $N$ denote the total number of negatives in the ground truth. In addition, for the outcome of the algorithm, if the algorithm outputs one edge that does exist in the ground truth, it scores a true positive ($TP$). If the algorithm outputs one edge that does not exist in the ground truth, it scores a false positive ($FP$).  The behavior of the algorithms is studied by using the receiver operating characteristic (ROC) curve \cite{marbach2012wisdom}, i.e. $TP$ rate ($TPR$) vs $FP$ rate ($FPR$) over different choices of their tuning parameters, where
\begin{equation*}
TPR = \frac{TP}{P}, \quad FPR = \frac{FP}{N},
\end{equation*}
which are further averaged over the number of Monte Carlo experiments. The tuning parameter for the BS algorithm and the iterative-EM BS algorithm is $\tau \in \{0,1,...,10\}$, while the tuning parameters of GLasso and the kernel-based GLasso are $\delta_j \in \{0,10,20,...,2000\}$ and $\beta_j=\{0.1,...,0.9\}$. To build ROC curves for the two GLasso estimators, $\delta_j$ and $\beta_j$ are kept the same for all MISO problems to reduce the number of tuning parameters.
The $(0,1)$ point in the ROC plot denotes the ideal performance without any error. Thus, the points on ROC curves of different methods can be compared based on their closeness to the $(0,1)$ point, i.e. computing $dis = \sqrt{FPR^2+(1-TPR)^2}$. A smaller $dis$ value implies a better performance.

We consider dynamic networks with $6$ nodes and three experiment conditions with different data length $N$ and model order $n$ are considered: $N=2000$ and $n=100$; $N=500$ and $n=100$; $N=50$ and $n=50$.  Note that in the final study, the number of the postulated unknown parameters in the algorithm is larger than the number of the measurements. For each experiment condition, $50$ different data-generating systems and thus independent data sets are randomly generated as follows. For each data-generating system, its topology is generated by assigning a discrete uniform distribution to the existence of each edge and then we assign a random transfer function to every existing edge by using \textit{drmodel} function in Matlab. The orders of generated $G_{ji}$ and $H_j$ are randomly selected from $2$ to $5$ with a uniform distribution. To guarantee a reasonable signal-to-noise ratio, $G_{ji}$ is further normalized by its own $l_2$ norm. Finally, the data of the resulting system is obtained by injecting Guassian noises with zero mean and $\sigma_j(t) =1$, for all $j$ and $t$.

For each data set, to initialize the hyperparameter vector for the EM algorithm, we set $\hat{\beta}_{ji}^{(0)} = 0.5$, $\hat{\lambda}_{ji}^{(0)} = 0.5$ for all modules and $\hat{\sigma}_j^{(0)} (t)$ is the same for all $j$ and $t$, which is drawn from a norm distribution with mean $1$ and standard deviation $0.2$.


The obtained ROC curves are summarized in Fig~\ref{figabc}. For the kernel-based GLasso, since $\beta_j =0.7$ typically provides the best performance, only the ROC curves corresponding to $\beta_j =0.7$ are shown. It can be found that in all tests, the two search algorithms perform better than the two GLasso estimators because the ROC curves of the search algorithms are closer to the $(0, 1)$ point for every value of $\tau$. To compare the performance of the iterative-EM BS and the BS algorithm, the following measure is
used:
\begin{equation*}
V=[\sum_{i=1}^{11} \frac{dis_{iter-EM BS,i} - dis_{BS,i}}{dis_{BS,i}}] \div 11 \times 100\%,
\end{equation*}
where $i$ denotes the $i$th value of $\tau$ in $\{0,1,...,10\}$. Given one value of $\tau$, one point on the ROC curve is correspondingly selected and thus $dis_{BS,i}$ can be calculated based on Fig~\ref{figabc}. Note that a positive
value of $V$ implies a worse performance of the iterative-EM algorithm. It can then be found that $V = -4\%$ when $N = 2000$, $V = 1\%$ when $N = 500$ and $V = 18\%$ when $N = 50$. Thus,  the iterative-EM BS algorithm performs better than the BS algorithm when
$N$ is large while it has worse performance when the sample size is relatively small. Intuitively, this can be explained by the fact that the iterative-EM algorithm relies more on the data because it adjusts the parameter prior given every different graph during the search procedure, leading to a larger error when the data length is limited. The computational speed of the iterative-EM algorithm is also around 10 times slower in this $6$-node example. Thus, it is suggested to use the BS algorithm when $N$ is small and the faster computation is preferred.
%

The performance of the algorithms is also compared when cross validation is employed for the two GLasso estimators while $\tau$ equals to the default value, i.e. $\tau = 0$, for the two BS algorithms. For the cross validation, the training data contains the data up to time $2(N + 1)/3$ and the data left is kept for validation. The tuning parameter that provides the smallest root-mean-square error in predicting the validation data is selected. Note that in this case, the tuning parameters of the two GLasso estimators are allowed to be different over the MISO problems. The final results contain one $(FPR, TPR)$ point for every algorithm and their distance to $(0, 1)$ is summarized in Table~\ref{my-label}.
\begin{table}[h]
\centering
\caption{Distance of the results of the algorithms to $(0, 1)$ with the cross-validated or the default tuning parameter}
\label{my-label}
\resizebox{.45\textwidth}{!}{%
\begin{tabular}{|l|l|l|l|l|}
\hline
 & BS & Iter-EM BS & GLasso  & K-GLasso  \\ \hline
$N=2000$ &$0.04$  & $0.04$ & $0.59$ & $0.64$ \\ \hline
$N=500$ & $0.07$ & $0.07$ & $0.37$ & $0.60$ \\ \hline
 $N=50$ & $0.20$ & $0.22$ & $0.52$ & $0.47$ \\ \hline
\end{tabular}}
\end{table}

No significant difference is observed between the BS and the iterative-EM BS algorithm while the two search algorithms outperform the two GLasso estimators due to their smaller distance to $(0,1)$. This is because the cross validation is designed for obtaining the tuning parameters corresponding to the best prediction performance, which typically leads to a model with more positives to improve the prediction. Instead, the Bayes factor typically favors simpler models, which may lead to a model with poorer prediction performance. This difference in the design purpose between BIC, which is an asymptotic approximation of the Bayes factor, and cross validation is also mentioned in  \cite{gelman2014understanding}. 

\section{Conclusion}
A Bayesian approach for topology identification of networks of transfer functions is explored. It uses the Bayes factor coupled with a forward-backward search algorithm. The Bayes factor is obtained by modeling the infinite impulse responses of the modules as Gaussian processes, where the hyperparameters of the Gaussian prior are estimated by the EM algorithm. Numerical results demonstrate the effectiveness of the algorithm, which  shows better performance compared to the group Lasso estimator.



\section*{APPENDIX}
\subsection{Proof of Proposition~\ref{pro:EM}}
Recall the notations defined in (\ref{eq:EM_sigma}). The proof contains two steps, including the E-step and the M-step of the EM algorithm.

\begin{proof}
\textbf{E-step}: Firstly, note that $\log P(\theta_j,w^N_j; \eta_j) = \log P(w^N_j| \theta_j ; \eta_j) + \log P(\theta_j ; \eta_j)$, where $\log P(w^N_j| \theta_j ; \eta_j)$ is the likelihood function given by the model and $\log P(\theta_j ; \eta_j)$ is the parameter prior of the full graph given by (\ref{eq:parameter prior}). Thus, it can be found that
\begin{align*}
\log P(\theta_j,w^N_j; \eta_j) =& \text{constant} - \frac{1}{2} \sum_{i=1}^L \log \det(\lambda_{ji}\bar{K}(\beta_{ji})) \\
& -\frac{1}{2} \log \det(\sigma_j^2I_N) -\frac{1}{2\sigma^2_j} (w^N_j)^Tw^N_j \\
& -\frac{1}{2}\theta_j^T \Sigma_j \theta_j + \frac{1}{\sigma^2_j} (w^N_j)^T A_j\theta_j  ,
\end{align*}
where $\Sigma_j$ is formulated as in (\ref{eq:EM_sigma}) given $\eta_j$.

$Q(\eta_j,\hat{\eta}_j^{(k)})$ can then be obtained by calculating the expectation of $\log P(\theta_j,w^N_j; \eta_j)$ over the posterior distribution of $\theta_j$ given the data and $\hat{\eta}^{(k)}$. Due to the Gaussian noise and the parameter prior (\ref{eq:parameter prior}), it follows that the posterior distribution of the parameter also has a Gaussian distribution as
\begin{equation*}
\theta_j | w_j \sim \mathcal{N}(\hat{C}^{(k)}_jw^N_j,{(\hat{\Sigma}_j^{(k)}})^{-1}).
\end{equation*}

Thus, the E-step can be finalized as
\begin{equation}
Q(\eta_j,\hat{\eta}_j^{(k)}) = Q_1(\sigma_j,\hat{\eta}_j^{(k)}) + \sum_{i=1}^L Q_2(\lambda_{ji},\beta_{ji},\hat{\eta}_j^{(k)})+ \text{constant}, \label{eq:EM_Qfunction}
\end{equation}
where
\begin{align}
Q_1(\sigma_j,\hat{\eta}_j^{(k)}) =& -\frac{1}{2} tr(\sigma_j^{-2}A^T_jA_j\hat{\Delta}^{(k)}_j)  -N \log \sigma_j \nonumber   \\
& +\frac{1}{\sigma_j^2}(w^N_j)^TA_j \hat{C}^{(k)}_jw_j \nonumber \\
 & - \frac{1}{2\sigma_j^2}(w^N_j)^Tw^N_j, \nonumber  \\
& = -N \log \sigma_j - \frac{1}{2\sigma^2_j} M^{(k)}, \label{eq:EM_Q1} \\
Q_2(\lambda_{ji},\beta_{ji},\hat{\eta}_j^{(k)}) = & -\frac{1}{2} \log \det[\lambda_{ji}\bar{K}(\beta_{ji})] \nonumber  \\
& - \frac{1}{2} tr[(\lambda_{ji}\bar{K}(\beta_{ji}))^{-1} \hat{\Delta}^{(k)}_j[i]]. \label{eq:EM_Q2}
\end{align}
where $M^{(k)}$ is formulated as shown in (\ref{eq:EM_sigma}).

It can be found that $Q$ is decomposed into two parts, including $Q_1$ as a function of $\sigma_j$ and $Q_2$ as a function of the parameters from the parameter prior. Thus, the optimization of $Q$ can be solved by considering $Q_1$ and $Q_2$ independently. The constant term in (\ref{eq:EM_Qfunction}) will be ignored because it does not influence the optimization result.

\textbf{M-step}:
It can be found that (\ref{eq:EM_Q1}) is maximized by (\ref{eq:EM_sigma}) assuming that $M^{(k)} >0$.

To maximize $Q_2(\lambda_{ji},\beta_{ji},\hat{\eta}^{(k)})$, set the derivative of (\ref{eq:EM_Q2}) over $\lambda_{ji}$ to be zero, which leads to the solution of $\lambda_{ji}$ as
\begin{equation}
\lambda_{ji}^* = \frac{1}{n}  tr[\bar{K}^{-1}(\beta_{ji}) \hat{\Delta}^{(k)}_j[i]], \label{eq:EM_lamda_pre}
\end{equation}
which is a function of $\beta_{ji}$. Plugging (\ref{eq:EM_lamda_pre}) back into (\ref{eq:EM_Q2}), one obtains that
\begin{align*}
Q_2(\lambda^*_{ji},\beta_{ji},\eta^{(k)}) =& -\frac{n}{2} \log [tr(\bar{K}^{-1}(\beta_{ji}) \hat{\Delta}^{(k)}_j[i])]\\
& -\frac{1}{2} \log \det\bar{K}(\beta_{ji})+\text{constant},
\end{align*}
which can be maximized by minimizing (\ref{eq:EM_beta}). After obtaining $\hat{\beta}^{(k+1)}_{ji}$, $\hat{\lambda}^{(k+1)}_{ji}$ can be found by (\ref{eq:EM_lamda}). Thus, $Q_1$, $Q_2$ have been optimized independently and M-step is proved.
\end{proof}

%
%
%
%
%

\bibliographystyle{IEEEtran} 
\bibliography{ReferByes}
\addtolength{\textheight}{-12cm}   

\end{document}